\documentclass{aastex}

\usepackage{emulateapj5}

\input{epsf}

\submitted{Accepted by ApJL on August 29, 2000}

\shorttitle{Formation of cuspy density profiles}

\shortauthors{Ewa L. \L okas and Yehuda Hoffman}

\begin{document}

\title{Formation of cuspy density profiles\\- a generic feature of
collisionless gravitational collapse}

\author{Ewa L. \L okas}

\affil{Copernicus Astronomical Center, Bartycka 18, 00--716 Warsaw, Poland}

\and

\author{Yehuda Hoffman}

\affil{Racah Institute of Physics, Hebrew University, Jerusalem 91904,
Israel}



\begin{abstract}

Using the formalism of the spherical infall model the structure of
collapsed and virialized dark halos is calculated for a variety of
scale-free initial conditions.  In spite of the scale-free
cosmological
nature of the problem, the collapse of individual objects is not
self-similar.  Unlike most of previous calculations the dynamics used
here relies only on adiabatic invariants and not on self-similarity.
The paper focuses on the structure of the innermost part of the
collapsed halos and addresses the problem of central density cusps.
The slopes of density profiles at 1\% of virial radius are calculated
for a variety of cosmological models and are found to vary with the
mass of the halos, power spectrum of density fluctuations and
$\Omega_0$. The inner slopes range between $r^{-2.3}$ and $r^{-2}$ with
the limiting case of $r^{-2}$ reached for the largest masses.  The steep
cusps found here correspond to the limiting case where all particles move
on radial orbits.  The introduction of angular momentum will make the
density profile shallower.  We expect this to resolve the discrepancy
found between the calculated profiles and the ones found in
high resolution $N$-body simulations, where the exponent ranges from
$-0.5$ to $-1.5$. The robust prediction here is that collisionless
gravitational collapse in an expanding universe is expected to form
density cups and not halos with a core structure.

\end{abstract}


\section{Introduction}

The structure of collapsed and virialized objects such as galaxies and
clusters poses a real challenge to our understanding of structure
formation in the universe.  The basic problem in this field
is that of the formation of `dark halos', namely the outcome of the
collisionless collapse in an expanding universe.  This was addressed
first by two seminal papers of Gunn \& Gott (1972) and Gunn
(1977), where the cosmological expansion and the role of adiabatic
invariance were first introduced in the context of the formation of
individual objects. The next step was made by Fillmore \& Goldreich (1984)
and Bertschinger (1985) who found analytical predictions for the density
of collapsed objects seeded by scale-free primordial initial perturbation
in a flat universe.  Hoffman \& Shaham (1985, HS) applied and modified
these solutions to realistic initial conditions in flat as well as open
Friedman models. Their basic result was that for an
Einstein-de Sitter universe and a primordial power spectrum $P(k) \propto
k^{n}$ the asymptotic solution for the density profile is $\rho \propto
r^\alpha$ where $\alpha=-3(3+n)/(4+n)$, i.e. higher $n$ yield steeper
profiles (similar trend was found for lower $\Omega_0$). This model was
improved in a series of papers focusing on the incorporation of the peak
formalism of Bardeen et al. (1986, BBKS) or weakly nonlinear corrections
and the refinement of the adiabatic invariance calculations (Ryden \& Gunn
1987; Hoffman 1988; Ryden 1988; Zaroubi \& Hoffman 1993; \L okas 1998,
2000).

The analytical studies were followed by numerical studies of the
collapse problem by the means of $N$-body simulations e.g. by Quinn,
Salmon, \& Zurek (1986) and Crone, Evrard, \& Richstone (1994) who
confirmed predictions of HS. Recently Navarro, Frenk \& White (1997, NFW)
established that the density profile of dark matter halos forming in
different cosmologies follows a universal form that steepens from $r^{-1}$
near the center of the halo to $r^{-3}$ at large distances. This result
was confirmed by Cole \& Lacey (1996), Huss, Jain, \& Steinmetz (1999a)
and others, although Kravtsov et al. (1998) obtain much shallower inner
profiles. Some recent very high resolution cosmological simulations
produce steeper density profiles, with inner slopes $r^{-1.5}$ (Fukushige
\& Makino 1997, 2000; Moore et al. 1998; Jing \& Suto 2000). It seems that
the outcome of a collisionless gravitational collapse is a cuspy density
profile rather then a core structure, although the exact slope of the cusp
is still under debate.

There have been a number of attempts to identify a mechanism
responsible for the formation of the cusp, mainly referring to the
merging formalism of Lacey \& Cole (1993).  Syer \& White (1998) and
Nusser \& Sheth (1999) claimed that the universal profile is a result
of hierarchical clustering by mergers of smaller halos into bigger
ones.  However, Moore et al. (1999) performed $N$-body simulations
with a cut-off in the power spectrum at small scales and also obtained
halos with cuspy density profiles. This proves that merging and
substructure does not play a critical role in the formation of density
cusps.

The aim of the present paper is to demonstrate that the density cusp is
indeed a generic feature of collisionless collapse in an expanding
universe that emerges also from the highly simplistic spherical infall
model. The model is presented in Section~2. The resulting inner density
profiles are described in Section~3 and general discussion follows in
Section~4.

\section{The model}

The spherical infall model is based on two ingredients, namely the
initial conditions and the dynamical model. Given the initial density
structure a dynamical model is needed to describe the spherical collapse
in an expanding universe. The dynamics of a given mass shell has two
phases, before and after shell crossing occurs.  The first one is a
trivial dynamics of an isolated shell in which energy is conserved.  The
other is a much more complicated process which, as shown by Fillmore \&
Goldreich (1984), can be described by considering the adiabatic
invariance of the motion. In the case of scale-free mass distribution, for
a given spherical shell, characterized by a maximum radius $r$
and the mass that is confined within it $m(r)$, $r m(r)$ is an adiabatic
invariant.  The mass enclosed by a shell is composed of the primordial
mass that was confined within it $m_{0}(r)$ and a secondary component of
the mass contributed by external shells spending some fraction of their
time within it, $m_{\rm add} (r)$. It follows that the maximum
radius shrinks in time by a factor $f = m_{0}(r)/[m_{0}(r) + m_{\rm
add} (r)]$.

However, self-similar solutions have very limited validity in  trying to
calculate the structure of actual dark halos, although most of earlier
work on the subject relied on self-similarity (with the notable exception
of Ryden 1988). We use here the spherical infall model of \L okas (2000),
which provides a generalization of that proposed by HS. This version of
the model used the generalized (not scale-free) statistically expected
initial density distribution around an overdense region and introduced a
boundary of the evolving protosystem by cutting off the distribution at
half inter-peak separation.

Here we further modify the model by assuming that the dark halos are
seeded by local density maxima (not just overdense regions) and their
initial structure is given by the peak statistics (BBKS) of the linear
density field smoothed with a Gaussian filter of scale $R$. We assume
scale-free initial density fluctuation spectra $P(k) \propto k^n$ and
normalize them to $\sigma_8 = 1$. We apply here the peak density profile
averaged over curvatures and orientations as given by equation (7.10) of
BBKS and use the half inter-peak cut-off scales as calculated in \L okas
(2000).

The initial density distribution has to be specified further by adopting
appropriate initial conditions. We assume that the starting point in the
evolution of every halo is given by the condition for the overdensity
within $R$, $\delta=\nu \sigma=0.1$ (a value small enough for the linear
theory to be valid), where $\nu$ is the height of the peak and $\sigma
\propto R^{-(n+3)/2}$ is the rms fluctuation at scale $R$. Once $R$ is
chosen, this condition yields the initial redshift.

Our purpose is to study the present properties of dark halos, therefore we
assume that the collapse time (as defined in the spherical top-hat model)
for all objects is the present epoch. This determines the mean density
inside the presently collapsing shell. Comparing this value with the
density distribution around the peak we determine the radius of this
shell (typically a few $R$) and define it to be the virial radius. The
present turn-around radius is obtained in a similar way. The mass of the
halo is a sum of the mass inside the virial radius and the mass
contributed by the shells with maximum radii between the virial radius and
the turn-around radius.

Once the initial density structure up to the present turn-around radius is
determined, it is straightforward to predict the fiducial mass
distribution with all shells at their turn-around radii. We then apply the
adiabatic invariance approach to calculate the motion of shells in such
distribution and the $m_{\rm add}$ contributed for each shell by shells of
larger maximum radii (see  \L okas 2000). In the case of
absence of self-similarity, we find that for our initial mass
distribution the adiabatic invariant $J\propto\int_{0}^{r'} v(r,r') {\rm
d} r = {\rm const}$ ($r'$ being the shell's apapsis) leads to a more
complicated relation between the final radius $r_{\rm f}$ and mass:
$g(r_{\rm f})/g(r)=f^{1/2}$, where $g(r)$ can be well approximated by
$r^{1/2}$ multiplied by a polynomial correction. The collapse factor
$F=r_{\rm f}/r$ has to be obtained by solving numerically this equation
and typically yields values of a few percent higher than $f$, with the
difference being larger for larger masses. An example of the difference
between $F$ and $f$ is shown in Figure~1. It turns out that the proper
calculation of the collapse factor in the case of departure from
self-similarity is crucial to the formation of central density cusps in
dark halos. Previous calculations, where the approach based on
self-similarity was used, did not reproduce the density cusps (del Popolo
et al. 2000).

\begin{center}
    \leavevmode
    \epsfxsize=7cm
    \epsfbox[50 50 340 310]{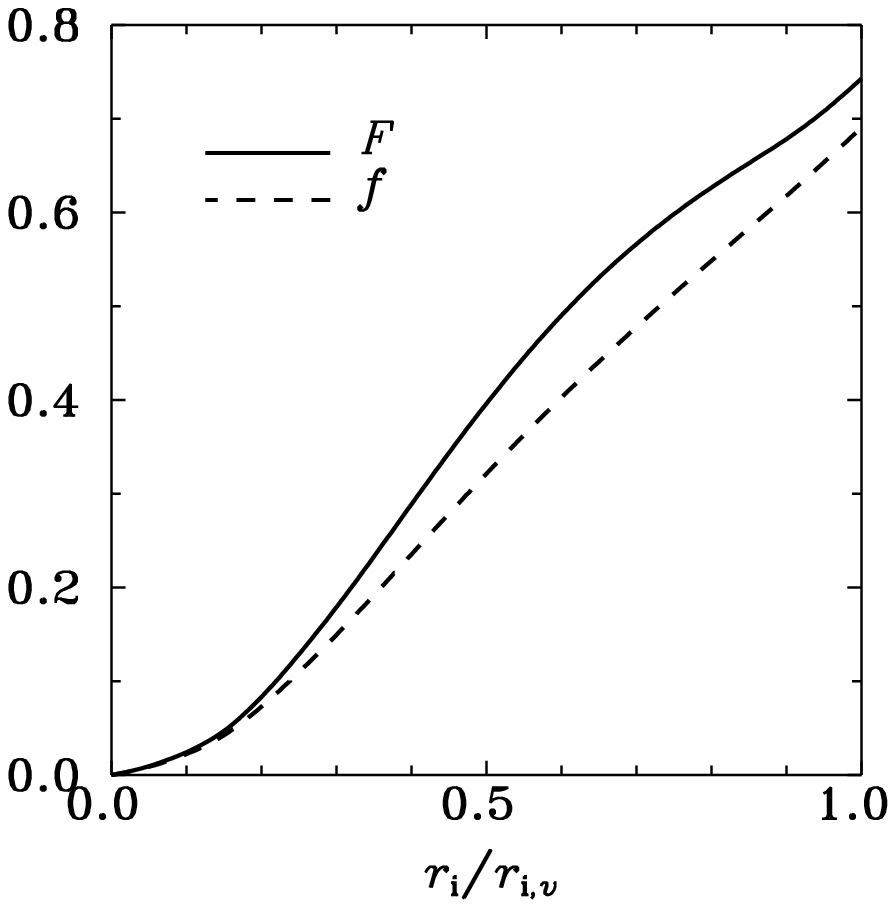}
\end{center}
    \figcaption{Collapse factors $F$ and $f$ as functions of
    initial radius (in units of the virial radius) for one particular
    case of the density profile of a halo of mass $M/M_{\ast} = 0.78$
    obtained in the case of $\Omega_0=1$, $n=-1$, $z_{\rm i}=100$ and
    $w=0.5$.}
\label{factor}

\vspace{0.1in}

We also generalized the improved spherical infall formalism to apply to the
open universe models. This is straightforward and involves only some
changes in the relation between the maximum radius and the initial radius
of the shell (HS), the rate of growth of linear density fluctuations and
the formula for the present age of the universe, which are all well known
(e.g. Padmanabhan 1993).

\section{Results}

In this section we present the results of application of the presented
model to the calculation of cusps, i.e. the slopes of the final density
profiles at 1\% of the virial radius.

The most uncertain part of the model is the determination the boundaries
of the collapsing halo.  As mentioned, it is assumed to correspond to the
half inter-peak separation. We adopt the same smooth radial
cutoff of exponential shape as in \L okas (2000) with a parameter $w$ (a
width of the cut-off filter in units of the smoothing scale $R$) as a
measure of its sharpness: the smaller the $w$ the sharper the cut-off. One
may argue that $w$ should be of the order of unity, since this is the
resolution associated with smoothing of the initial density field. However,
the concentrations of final density profiles depend strongly on $w$ which
significantly reduces the predictive power of the spherical infall model.

Given the uncertainty in determining the cut-off, we check how the slopes
of cusps depend on $w$. Figure~2 shows the effective slopes $\alpha$ of
the final density profiles, $\rho \propto r^\alpha$, at 1\% of the virial
radius calculated with three different values of $w=0.5, 1$ and $2$ in
the case of $\Omega_0=1$, $\nu=3$ and two values of the spectral index $n$.
The mass dependence comes from different smoothing scales corresponding
to different initial redshifts. The plotted range of masses was obtained
with initial redshifts in the range $1500 < z_{\rm i} < 100$. The masses
were expressed in terms of the so-called present nonlinear mass $M_{\ast}$
which depends on the cosmological model and reflects how advanced is the
formation of nonlinear objects at present (see NFW; \L okas 2000).

\begin{center}
    \leavevmode
    \epsfxsize=7cm
    \epsfbox[50 50 340 310]{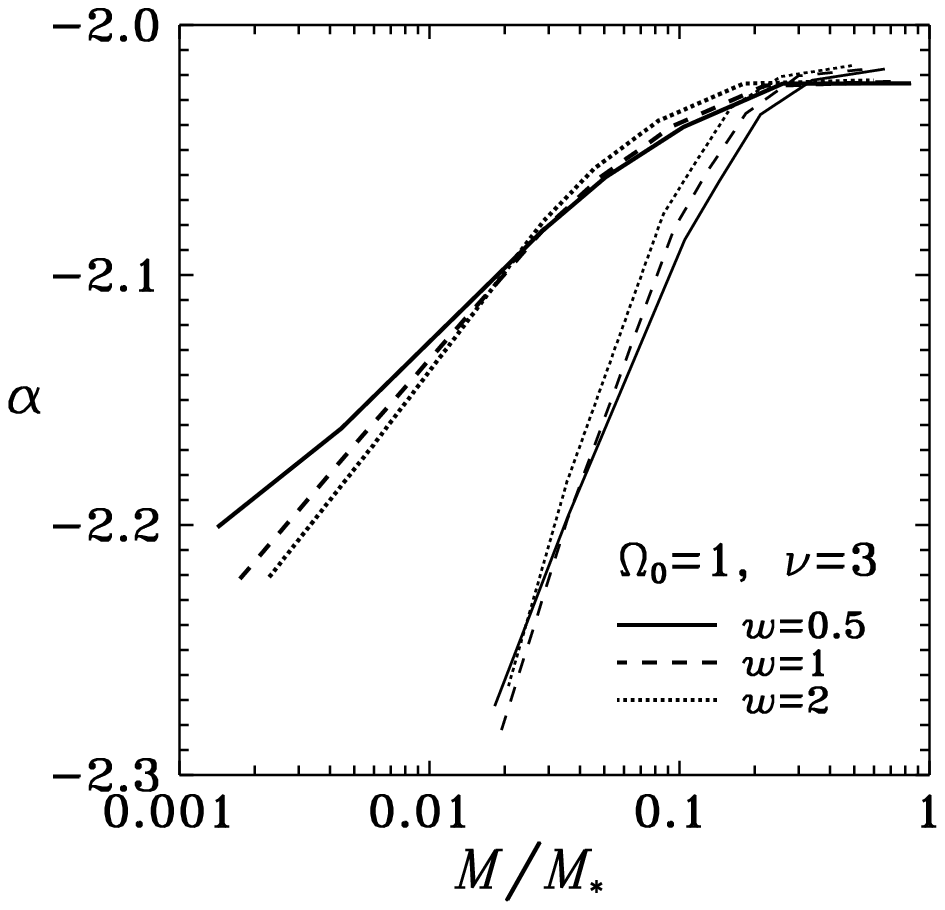}
\end{center}
    \figcaption{Slopes of the inner profiles of halos as a function of
    their mass obtained with different parameters $w$ for $\Omega_0=1$,
    $\nu=3$ and two spectral indices: $n=-1$ (thicker lines) and $n=0$
    (thinner lines).}
\label{cuspw}

\vspace{0.1in}

Fortunately, we find that the range of cusp profiles obtained for
different masses depends very weakly on $w$. The slopes of the inner
profiles vary between $\alpha=-2.3$ and $\alpha=-2$ when going from the
smallest to the largest masses, being somewhat steeper for $n=0$ than for
$n=-1$, in agreement with the original prediction of HS.

Next, we explore the dependence of the inner profiles on the initial
conditions taking into account different heights of the peak in the
initial density field, $\nu$, that gives rise to a bound object.  Our
condition $\nu \sigma=0.1$ leads to degeneracy in
this picture of structure formation: when $\nu$ is specified this
condition enables us to relate uniquely the mass of the halo to the
smoothing scale $R$.  If we let $\nu$ vary, halos of the same mass can
result from assumptions of different $(\nu, R)$ pairs.

\begin{center}
    \leavevmode
    \epsfxsize=7cm
    \epsfbox[50 50 340 310]{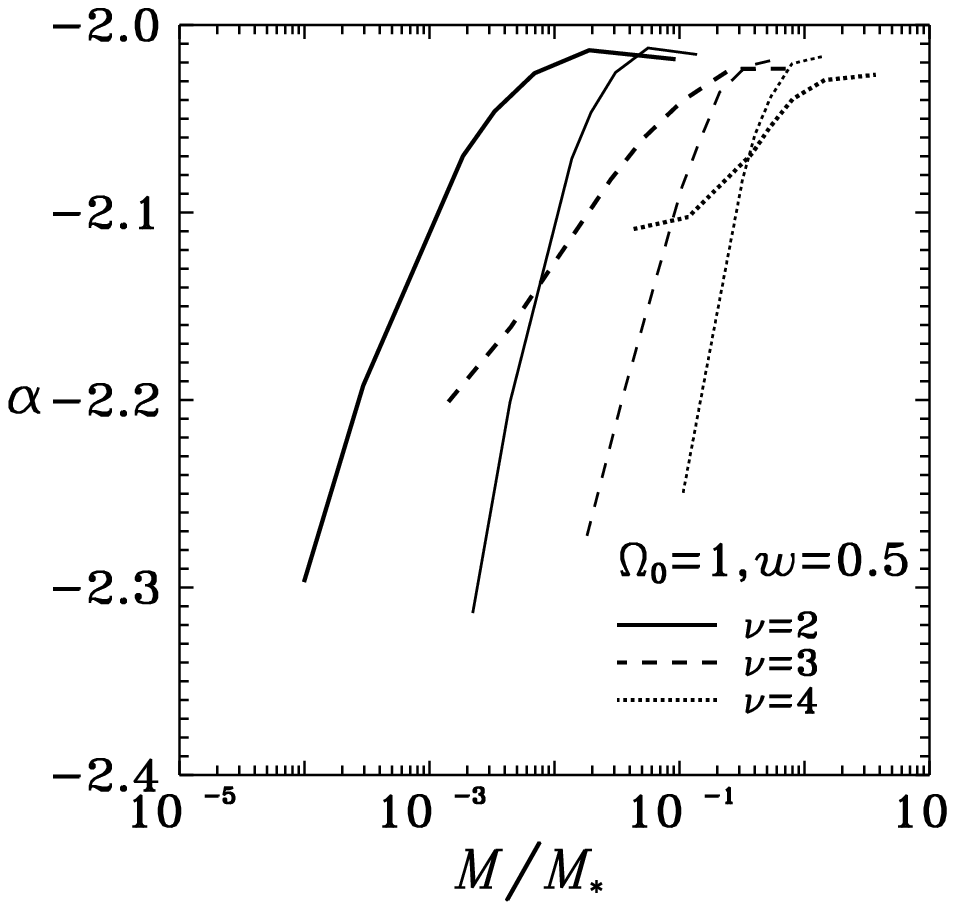}
\end{center}
    \figcaption{Slopes of the inner profiles of halos as a function of
    their mass for different heights of the peak $\nu$, for $\Omega_0=1$
    and two spectral indices: $n=-1$ (thicker lines) and $n=0$ (thinner
    lines).}
\label{cuspa}

\vspace{0.1in}

We test the dependence of the inner profiles on the choice of $\nu$ by
calculating the cusps for $\nu=2$ and $4$ in addition the results for
$\nu=3$ presented above.  We accordingly adjust the scale of the
cut-off in the initial density distribution (it is smaller for lower
peaks, see \L okas 2000), but choose only the width $w=0.5$, since, as
we proved with Figure~2, the results depend weakly on its value.
The predictions of the inner profiles for different $\nu$ are shown in
Figure~3. The trend of steeper cusps for lower masses seen in
Figure~2 is preserved for all values of $\nu$, but smaller $\nu$
(with the rest of initial conditions unchanged) typically produce smaller
masses so the curves in Figure~3 are shifted accordingly.
Figure~4 shows how the results depend on $\Omega_0$. Again, our
predictions agree with the trend of steeper profiles for lower $\Omega_0$
found by HS.

\begin{center}
    \leavevmode
    \epsfxsize=7cm
    \epsfbox[50 50 340 310]{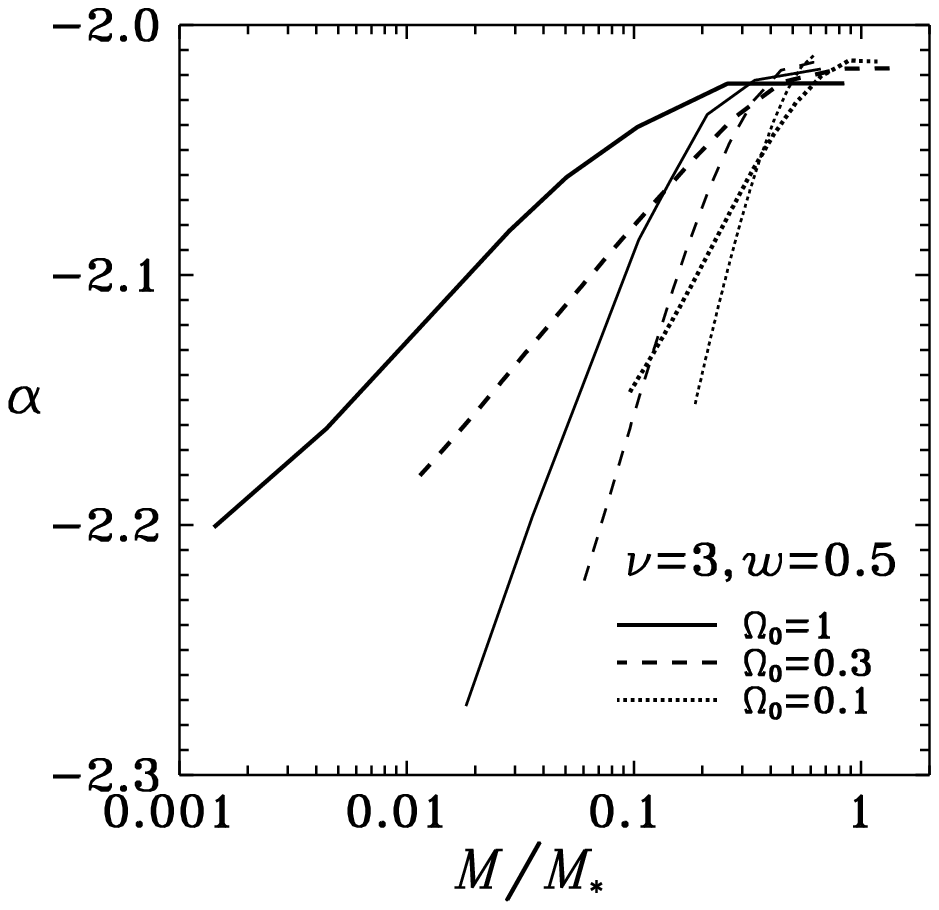}
\end{center}
    \figcaption{Slopes of the inner profiles of halos as a function of
    their mass for different values of $\Omega_0$ (with height of the peak
    $\nu=3$) and for two spectral indices: $n=-1$ (thicker lines) and $n=0$
    (thinner lines).}
\label{cuom}

\vspace{0.1in}

\section{Discussion}

In the limit of large masses we reproduce the result of Fillmore \&
Goldreich (1984) and Zaroubi \& Hoffman (1993) who found that in the case
of scale-free initial density distribution if this distribution is flat
enough, the final density profile will be $\rho \propto r^{-2}$.
This result can be interpreted in terms of the shape of the initial
density distribution: the profiles of halos of larger mass originate
from shells that initially were quite close to the peak, so their
cumulative density distributions were rather flat.

When comparing our results to those of $N$-body simulations one must be
careful about the meaning of the slope of the cusp. In the
case of NFW profile the limiting inner slope of their fitting formula
when $r \rightarrow 0$ is $-1$, but this is not equal to the effective
slope at $0.01 r_v$, which instead is $\alpha = -(1+0.03 c)/(1+0.01 c)$,
where $c$ is the concentration, i.e. the ratio of the virial radius to the
scale radius, $r_{\rm s}$ (the radius at which the slope is $-2$).
Clearly, $r_{\rm s}=0.01 r_v$ or $c=100$ is enough to have $\alpha=-2$.
For example, in the range of masses shown in Figure~2 for $n=-1$ the
effective inner slopes calculated with the above formula (using the
description NFW provide for the dependence of the concentration on mass)
change between $\alpha = -2.2$ for the smallest mass and $\alpha=-1.3$ for
the largest, that is they are quite far from the limiting slope of $-1$.
Comparison of these values with those predicted by our model shows that
reasonable agreement is obtained for smaller masses, while for larger
masses NFW predict much shallower cusps. Unfortunately, this conclusion is
based on an extrapolation of the $N$-body results to the range of small
masses, because in the simulations the halos are typically more massive
than those we obtain. A better agreement is expected with those authors
of $N$-body simulations, who claim the inner slope to be closer to $-1.5$.

In general, however, the spherical infall model predicts inner slopes
of the profiles that are steeper than those observed in $N$-body
simulations.  This result is expected, given the simplistic nature of
the model where radial orbits are assumed.  This assumption maximizes
the secondary mass added to the original mass enclosed within a given
shell, and therefore reinforces contraction which steepens
the density profile.  In reality, angular momentum prevents
a given shell from pulling inside some of the inner shells, which
results in weaker adiabatic contraction of these shells.  This effect
was noticed before by Ryden (1988), Sikivie, Tkachev, \& Wang
(1997), Avila-Reese, Firmani \& Hernandez (1998) and Huss, Jain, \&
Steinmetz (1999b) who found that addition of angular momentum or
tangential velocity dispersion can flatten the density profile even to
$\rho \propto r^{-1}$ instead of the $\rho \propto r^{-2}$ expected for
radial infall.

Comparisons of the theoretically predicted density profiles with
observations have recently put the CDM-based scenario of structure
formation in trouble (Moore et al. 1999).  A detailed comparison is
beyond the scope of the present paper, but we conclude with a few
comments. The strongest case against the claim that galaxies form
via collisionless gravitational collapse is that of the core
structure (or shallow cusp) of low surface brightness (LSB) galaxies
(Kravtsov et al. 1998; Moore et al. 1999). However, van den Bosch et al.
(1999) claim that in most cases observations of LSBs have low resolution
and can place little constraints on the inner shapes of density profiles.
In the cases with high enough resolution they find agreement with NFW.
Anyway, LSBs are more angular momentum dominated, compared to normal
galaxies with the same luminosity (McGaugh \& de Blok 1999), and therefore
we argue that these objects should typically have shallower density
cusps. The strongest claims against CDM based on the rotation curves of
dwarf galaxies have also been recently challenged (van den Bosch \& Swaters
2000) due to observational uncertainties. It is interesting to note here
that steep inner profiles similar to NFW are roughly consistent with the
observations of elliptical galaxies (\L okas \& Mamon 2000). Thus we
conclude that the paradigm of collisionless gravitational collapse should
not yet be dismissed.

\acknowledgments

E. de Blok is gratefully acknowledged for educating us on LSB
galaxies. We thank our referee, Adi Nusser, for comments which helped to
improve the paper. EL\L \ kindly acknowledges the hospitality of The
Hebrew University where this project was started. This work was supported
by the Polish KBN grants 2P03D00813 and 2P03D02319 as well as by the Israel
Science Foundation grant 103/98.


\begin{thebibliography}{}


\bibitem[]{afh} Avila-Reese, V., Firmani, C., \& Hernandez, X. 1998, ApJ,
    505, 37
\bibitem[]{bbks} Bardeen, J. M., Bond, J. R., Kaiser, N., \& Szalay,
    A. S. 1986, ApJ, 304, 15 (BBKS)
\bibitem[]{ber} Bertschinger, E. 1985, ApJS, 58, 39
\bibitem[]{cl} Cole, S., \& Lacey, C. 1996, MNRAS, 281, 716
\bibitem[]{cer} Crone, M. M., Evrard, A. E., \& Richstone, D. O. 1994, ApJ,
    434, 402
\bibitem[]{pgrs} del Popolo, A., Gambera, M., Recami, E., \& Spedicato, E.
    2000, A\&A, 353, 427
\bibitem[]{fg} Fillmore, J. A., \& Goldreich, P. 1984, 281, 1
\bibitem[]{fm} Fukushige, T., \& Makino, J. 1997, ApJ, 477, L9
\bibitem[]{fm1} Fukushige, T., \& Makino, J. 2000, submitted to ApJ,
    astro-ph/0008104
\bibitem[]{g} Gunn, J. E. 1977,  ApJ,  218, 592
\bibitem[]{gg} Gunn, J. E., \& Gott, J. R. 1972, ApJ, 176, 1
\bibitem[]{ho} Hoffman, Y. 1988, ApJ, 328, 489
\bibitem[]{hs} Hoffman, Y., \& Shaham, J. 1985, ApJ, 297, 16 (HS)
\bibitem[]{hjs1} Huss, A., Jain, B., \& Steinmetz, M. 1999a, MNRAS, 308,
    1011
\bibitem[]{hjs2} Huss, A., Jain, B., \& Steinmetz, M. 1999b, ApJ, 517, 64
\bibitem[]{js} Jing, Y. P., \& Suto, Y. 2000, ApJ, 529, L69
\bibitem[]{kkbp} Kravtsov A. V., Klypin A. A., Bullock J. S., Primack J.
    R. 1998, ApJ, 502, 48
\bibitem[]{lc} Lacey, C., \& Cole, S. 1993, MNRAS, 262, 627
\bibitem[]{lk} \L okas, E. L. 1998, MNRAS, 296, 491
\bibitem[]{lo} \L okas, E. L. 2000, MNRAS, 311, 423
\bibitem[]{lm} \L okas, E. L., \& Mamon, G. A. 2000, submitted to MNRAS,
    astro-ph/0002395
\bibitem[]{mb} McGaugh, S. S., \& de Blok, W. J. G. 1998, ApJ, 499, 41
\bibitem[]{mgqsl} Moore, B., Governato, F., Quinn, T., Stadel, J., \&
    Lake, G. 1998, ApJ, 499, L5
\bibitem[]{mq} Moore, B., Quinn, T., Governato, F., Stadel, J., \& Lake,
    G. 1999, MNRAS, 310, 1147
\bibitem[]{nfw} Navarro, J. F., Frenk, C. S., \& White, S. D. M. 1997,
    ApJ, 490, 493 (NFW)
\bibitem[]{ns} Nusser, A., \& Sheth, R. K. 1999, MNRAS, 303, 685
\bibitem[]{pa} Padmanabhan, T. 1993, Structure Formation in the Universe
    (Cambridge: Cambridge Univ. Press)
\bibitem[]{qsz} Quinn, P. J., Salmon, J. K., \& Zurek, W. H. 1986,  Nature,
    322, 329
\bibitem[]{stw} Sikivie, P., Tkachev, I. I., \& Wang, Y. 1997, Phys. Rev.
    D, 56, 1863
\bibitem[]{sw} Syer, D., \& White, S. D. M. 1998, MNRAS, 293, 337
\bibitem[]{r} Ryden, B. S. 1988, ApJ, 333, 78
\bibitem[]{rg} Ryden, B. S., \& Gunn, J. E. 1987, ApJ, 318, 15
\bibitem[]{brdb} van den Bosch, F. C., Robertson, B. E., Dalcanton, J. J.,
    \& de Blok, W. J. G. 1999, submitted to AJ, astro-ph/9911372
\bibitem[]{bs} van den Bosch, F. C., \& Swaters, R. A. 2000, submitted
    to AJ, astro-ph/0006048
\bibitem[]{zh} Zaroubi, S., \& Hoffman, Y. 1993, ApJ, 416, 410


\end{thebibliography}
\end{document}